\documentstyle[12pt]{article} \textwidth=6in \oddsidemargin=0in
\begin{document}
\newcommand{\dl}{\delta} \newcommand{\iy}{\infty} \newcommand{\La}{\Lambda}
\newcommand{\pl}{\partial} \newcommand{\noi}{\noindent}
\renewcommand{\sp}{\vskip2ex}
\newcommand{\bq}{\begin{equation}} \newcommand{\eq}{\end{equation}}
\newcommand{\ra}{\rightarrow} \newcommand{\al}{\alpha} \newcommand{\th}{\theta}
\newcommand{\bR}{{\bf R}}  \newcommand{\bC}{{\bf C}} \newcommand{\bZ}{{\bf Z}}
\newcommand{\nm}{\parallel} \newcommand{\la}{\lambda} \newcommand{\ph}{\varphi}
\newcommand{\om}{\omega} \newcommand{\inv}{^{-1}}\newcommand{\ve}{\varepsilon}
\newcommand{\Ga}{\Gamma} \newcommand{\hf}{{1\over 2}} \newcommand{\cH}{\cal H}
\newcommand{\ch}{\raisebox{.4ex}{$\chi$}}
\newcommand{\ov}{\over} \newcommand{\Om}{\Omega}
\newcommand{\ba}{\left(\begin{array}{cc}}
\newcommand{\ea}{\end{array}\right)} \newcommand{\cM}{{\cal M}}
\newcommand{\be}{\beta} \newcommand{\uv}{\Big({u\ov v}\Big)^{\be}}
\renewcommand{\sl}{s_{\la}}
\newcommand{\ub}{\,u^{\be}} \newcommand{\um}{\,u^{-\be}}
\newcommand{\tl}{\widetilde} \newcommand{\ds}{\displaystyle}
\newcommand{\ssl}{\textstyle}
\newcommand{\arccosh}{{\rm arccosh}}

\begin{center}{ \large\bf  Asymptotics of a Class of Solutions}\end{center}
\begin{center}{\large\bf to the Cylindrical Toda Equations}\end{center}
\sp\begin{center}{{\bf Craig A. Tracy}\\
{\it Department of Mathematics and Institute of Theoretical Dynamics\\
University of California, Davis, CA 95616, USA\\
e-mail address: tracy@itd.ucdavis.edu}}\end{center}
\begin{center}{{\bf Harold Widom}\\
{\it Department of Mathematics\\
University of California, Santa Cruz, CA 95064, USA\\
e-mail address: widom@math.ucsc.edu}}\end{center}\sp

\renewcommand{\theequation}{1.\arabic{equation}}
\begin{abstract}
The small $t$ asymptotics of a class of solutions to
the $2D$ cylindrical Toda equations is computed.  The solutions, $q_k(t)$,
have the representation
\[ q_k(t)=\log\,\det\,(I-\la\,K_k)-\log\,\det\,(I-\la\,K_{k-1}) \]
where $K_k$ are  integral operators.  This class includes
the $n$-periodic cylindrical Toda equations.  For $n=2$ our results
reduce to the previously computed asymptotics of the $2D$ radial sinh-Gordon
equation and for $n=3$ (and with an additional symmetry constraint) they reduce
to earlier results for the
radial Bullough-Dodd equation. Both of these special cases are examples of
Painlev{\'e} III and have arisen in various applications.
The asymptotics of $q_k(t)$ are derived by  computing the small $t$
asymptotics
\[\det\,(I-\la\,K_k)\sim b_k\,\Big({t\ov n}\Big)^{a_k} \]
where explicit formulas are given for the quantities $a_k$ and $b_k$.  The
method
consists of showing that the resolvent operator of $K_k$ has an approximation
in terms of resolvents of certain Wiener-Hopf operators, for which there
are explicit integral formulas.
\end{abstract}

\noi{\bf 1. Introduction}\sp
We consider here solutions of the cylindrical Toda equations
\bq
q_k''(t)+t^{-1}q_k'(t)=4\,(e^{q_k(t)-q_{k-1}(t)}-e^{q_{k+1}(t)-q_k(t)}),\qquad
k\in{\bZ},
\label{teq}\eq
satisfying the periodicity conditions $q_{k+n}=q_k$. The integer
$n$ is arbitrary but fixed. It follows from results in \cite{W2} that solutions
valid for
all $t>0$ are given by
\bq q_k(t)=\log\,\det\,(I-\la\,K_k)-\log\,\det\,(I-\la\,K_{k-1}),\label{qk}\eq
where $K_k$ is the integral operator on ${\bf R}^+$ with kernel
\bq\sum_{\om}\om^k\,c_{\om}{e^{-t[(1-\om)u+(1-\om\inv)u\inv]}\over-\om
u+v},\label
{Gkern}\eq
$\om$ running over the $n$th roots of unity other than 1.

In the case $n=2$ we have $q_{k+1}=-q_k$ and (\ref{teq}) becomes, with $q$
equal to
either $q_k$,
\[q''(t)+t^{-1}q'(t)=8\,\sinh2q(t),\]
which can be reduced to a particular case of the Painlev\'e III equation. The
connection with Fredholm
determinants was discovered by McCoy, Tracy and Wu \cite{MTW}, and in the same
paper
the asymptotics as $t\ra0$ of these solutions $q(t)$ were determined. (Note
that all
asymptotics as $t\ra\iy$ are trivial.) The asymptotics as $t\ra0$ of
 $\det\,(I-\la^2K_0^2)=\det\,(I-\la\,K_0)\,\det\,(I+\la\,K_0)$ were determined
in \cite{T}.
(See also \cite{BT}, where the asymptotics were found for a family of
kernels including this one as a special case.)  The asymptotics
of $\det\,(I-\la\,K_0)$ itself were stated without proof in \cite{Z}.

A class of periodic cylindrical Toda equations arises in thermodynamic
Bethe Ansatz considerations \cite{C}. There the additional constraint
$q_{-k-1}=-q_k$
is imposed. The solutions (\ref{qk}) satisfy this constraint as
long as the coefficients $c_{\om}$ satisfy $c_{\om}=-\om^3\,c_{\om\inv}$. (This
follows from
the fact that $\det\,(I-K_k)=\det\,(I-K_{-k-2})$ in this case, which is proved
by applying the change of variable $u\ra u\inv$.) The case $n=3$ of this gives
the
cylindrical Bullough-Dodd equation ($q=q_3$ now)
\[q''(t)+t^{-1}q'(t)=4\,(e^{2q(t)}-e^{-q(t)}),\]
which can be reduced to another special case of Painlev\'e III. Asymptotics of
a class of solutions to
$P_{III}$ including this one were announced in \cite{K}.

This paper is devoted to the determination of
the asymptotics of the quantities $\det\,(I-\la\,K_k)$ in the general case,
under
the condition stated below. (In the final sections we shall compare our results
in
the cases $n=2$ and $3$ with those cited above.) We write $K$ for $K_0$ and
consider at first only the asymptotics of \break
$\det\,(I-\la\,K)$. This is no loss of generality since $K_k$ is obtained from
$K$ upon
replacing the coefficients $c_{\om}$ by $\om^k\,c_{\om}$. The problem reduces
to the
asymptotics of $\int_0^{\iy}R(u,u;\la)\,du$ as $t\ra 0$, where $R(u,v;\la)$ is
the resolvent
kernel of $K$, the kernel of  $K\,(I-\la\,K)\inv$. Using operator techniques,
we
show that $R(u,u;\la)$ is well-approximated on $[1,\iy]$ by the corresponding
function when the exponentials in (\ref{Gkern}) are replaced by
$e^{-t(1-\om)u}$
and on $[0,1]$ by the corresponding function when the exponentials are replaced
by
$e^{-t(1-\om\inv)u\inv}$. (Actually the kernels have to be modified first
by multiplying by factors $(u/v)^{\be}$ with $\be$ depending on $\la$.) We
shall show that
after these replacements we obtain operators
which can be transformed into Wiener-Hopf operators, whose resolvent kernels
have
explicit integral representations. By these means the problem becomes that of
determining
the asymptotics of certain integrals. This is achieved by contour-shifting, and
we find
in the end that as $t\ra 0$
\bq\det\,(I-\la\,K)\sim b\,\Big({t\ov n}\Big)^a\label{detform}\eq
with $a$ and $b$ constants given explicitly in terms of certain zeros
of the function
\[ h(s):=\sin\,\pi s-\la\,\pi\,\sum_{\om} c_{\om}\,(-\om)^{s-1}.\]
These are the values at $\la$ of those zeros which equal $1,\cdots,n$ when
$\la=0$.

To state the result precisely, we denote by $\al_k=\al_k(\la)\ (k\in\bZ)$ the
zeros
of this function
indexed so that $\al_k(0)=k$. The zeros depend analytically on $\la$ as long as
they are
unequal, and when $\la=0$ they are the
integers. We derive the asymptotics (\ref{detform}) under the
assumption that there is a path in the complex plane $\bC$ running from $0$ to
$\la$ such that everywhere on the path
\bq \Re\,\al_0<\Re\,\al_1,\ \ \Re\,\al_0<1,\ \ \Re\,\al_1>0,\label{01cond}\eq
and no zero lies in the strip $\Re\,\al_0 <\Re\,s<\Re\,\al_1 $.
With this assumption
the constants $a$ and $b$ are given by the formulas
\[a={1\over n}\sum_{\al}\,\al^2-{(n+1)(2n+1)\over6},\]
\[b={\prod_{|j|<n}G({j\over n}+1)^{n-|j|}\over\prod_{\al,\al'}G({\al-\al'\over
n}+1)},\]
where $\al$ and $\al'$ run over the set $\{\al_1(\la),\cdots,\al_n(\la)\}$ and
$G$ denotes
the Barnes $G$-function \cite{B}.

 From these formulas we obtain the asymptotics of the solutions (\ref{qk}). The
requirement
now is that everywhere on a path from 0 to $\la$ we have for all $k$
\[\Re\,\al_k<\Re\,\al_{k+1} ,\ \ k-1<\Re\,\al_k <k+1.\]
If this holds then
\[q_k(t)=A\log\Big({t\ov n}\Big)+\log B+o(1),\]
where for $k=1,\cdots, n$ the constants $A$ and $B$ are given by
\[A=2\,(\al_k-k),\ \ B=\prod_{1\leq j<k}{\Ga({\al_j-\al_k\over n}+1)
\ov\Ga({\al_k-\al_j\over n})}\prod_{k<j\leq n}{\Ga({\al_j-\al_k\over n})\ov
\Ga({\al_k-\al_j\over n}+1)},\]
and for other values of $k$ are given by periodicity.

As for the correct range of validity of the formulas, we conjecture that it is
enough that
$\Re\,\al_k <\Re\,\al_{k+1}$ for all $k$
for some path from 0 to $\la$, and that the extra condition $k-1<\Re\,\al_k
<k+1$ is
automatically satisfied then. In the cases which we consider in detail this is
so and we
obtain the correct range of validity. Another way of stating the condition is
as follows.
Define $\La$ to be the complement of
\bq\{\la: \Re\,\al_k=\Re\,\al_{k+1}\ \mbox{for some}\ k\}.\label{Lac}\eq
Then the region of validity should be the connected component of $\La$
containing $\la=0$.
The region for which we prove the result is the largest connected subset of
this set in
which the extra condition holds.\sp

\noi{\bf Remark}. It is shown in \cite{W2} that the more general class of
kernels
\[\int_{\Om}\om^k\,{e^{-t[(1-\om)u+(1-\om\inv)u\inv]}\over-\om
u+v}\,d\rho(\om),\]
gives a solution to the the cylindrical Toda equations by the same formulas.
Here
$\rho$ can be any finite complex measure supported on a compact subset $\Om$
of
\[\{\om\in\bC:\,\Re\,\om<1,\ \Re\,\om\inv<1\}.\]
This assures that the operator is trace class. In case $\Om$ is the set of
$n$th roots of
unity other than $1$ the condition is satisfied and the solution will clearly
be
$n$-periodic. We shall actually do everything in this more general case and we
find
asymptotic formulas of the form (\ref{detform}) for the corresponding
determinants, with
the constants $a$ and $b$ being given by integral formulas involving the
function $h(s)$ now defined by
\bq h(s):=\sin\,\pi s-\la\,\pi\,\int_{\Om}
(-\om)^{s-1}\,d\rho(\om).\label{h}\eq
In the periodic case $h(s)$ is itself periodic and the integrals are
expressible in terms of its
zeros in a strip of width $n$. This is why the result here is so explicit.
The requirements on the $\al_k$ stated above now refer to this function,
and we assume throughout that they are satisfied.

\setcounter{equation}{0}\renewcommand{\theequation}{2.\arabic{equation}}
\pagebreak
\noi{\bf 2. The approximating operators}\sp

Recall that $K$ is the operator on $L_2(\bR^+)$ with kernel
\[K(u,v)=\int_{\Om}{e^{-t[(1-\om)u+(1-\om\inv)u\inv]}\over-\om
u+v}\,d\rho(\om).\]
We denote by $R(u,v;\la)$ the resolvent kernel of $K$, the kernel of
$R_{\la}:=K\,(I-\la\,K)\inv$. It is well-known that
\[-{d\over d\la}\log\,\det\,(I-\la\,K)=\int_0^{\iy}R(u,u;\la)\,du,\]
the trace of the operator $R_{\la}$. Hence
\bq\log\,\det\,(I-\la\,K)=
-\int_0^{\la}\,\int_0^{\iy}R(u,u;\mu)\,du\,d\mu.\label{detrep}\eq

In this section we are going to find a good approximation to the integral
\linebreak
$\int_0^{\iy}R(u,u;\la)\,du$ when $\la$ satisfies the condition stated in the
Introduction.
(Afterwards we shall replace $\la$ by $\mu$ and integrate with respect to $\mu$
over
the path from 0 to $\la$ throughout which (\ref{01cond}) holds.)
We begin with an observation. If we multiply the kernel $K(u,v)$ above by
$(u/v)^{\be}$ for any $\be$ then the resulting kernel still represents a
bounded
operator on $L_2(\bR^+)$ because of the decay of the exponential factor at 0
and $\iy$ and,
although the resolvent kernel changes, its value on the diagonal $u=v$ does
not.
We are going to find approximations to the resolvent kernels for these
modified operators, and precisely which $\be$ we take depends on $\la$.

Here is how
we choose it. It follows from our main assumption that for each $\la$ there
there exists $\sl\in (0,1)$ such that the function $h(s)$ given by (\ref{h})
has
no zeros on the line $\Re\,s=\sl$. (In fact the assumption guarantees that
$\sl$ can be
chosen to vary continuously with $\la$.) With this $\sl$ we set $\be=\hf-\sl$.
Notice that $|\be|<\hf$, a fact we shall need in order to apply our
approximation argument.

We write the kernel as
\bq K_t(u,v)=\uv\int_{\Om}{e^{-t[(1-\om)u+(1-\om\inv)u\inv]}\over-\om
u+v}\,d\rho(\om)
\label{Kkern}\eq
and denote the operator itself by $K_t$. We do not display the dependence
on $\be$, which is fixed for now, but use the subscript $t$ to
help the reader distinguish those operators that depend on $t$ from those that
don't. Both
kinds will arise; the former have the subscript $t$ and the latter will not.

The exponential in (\ref{Kkern}) is the product
$e^{-t(1-\om)u}\,e^{-t(1-\om\inv)u\inv}$.
For $u\geq1$ the second factor is uniformly close to 1 when $t$ is small while
for $u\leq1$
the first factor
is uniformly close to 1. This suggests that the operators $K_t^{\pm}$ with
kernels
\[K_t^+(u,v):=\uv\int_{\Om}{e^{-t(1-\om)u}\over -\om u+v}\,d\rho(\om),\quad
K_t^-(u,v):=\uv\int_{\Om}{e^{-t(1-\om\inv)u\inv}\over -\om u+v}\,d\rho(\om),\]
should in some sense approximate $K_t$ on $u\geq1,\ u\leq1$, respectively, and
therefore
the resolvent kernels of these operators should approximate the resolvent
kernel of $K_t$
on these
intervals. We shall
show that this is so, and that if $R_t^{\pm}(u,v;\la)$ denote
the resolvent kernels of $K_t^{\pm}$, also on $\bR^+$, then
\bq\begin{array}{c}
\displaystyle\int_0^1R(u,u;\la)\,du=\int_0^1R_t^-(u,u;\la)\,du+o(1),\\ \\
\displaystyle\int_1^{\iy}R(u,u;\la)\,du=
\int_1^{\iy}R_t^+(u,u;\la)\,du+o(1)\end{array}
\label{resest}\eq
as $t\ra0$. We denote by $P^+$
multiplication by the characteristic function of $(1,\iy)$ and by $P^-$
multiplication by the characteristic function of $(0,1)$. We shall use the
notation $o_1(\la)$ to denote any family of operators whose trace norms are at
most $|\la|$
times a function of $t$ which is $o(1)$ as $t\ra0$. (The subscript 1 refers to
the trace
norm. We shall also use the obvious notation $o_1(1)$ later on.) The main
approximation
statement will be
\bq
P^{\pm}\,(I-\la\,K_t)\inv\,P^{\pm}
=P^{\pm}\,(I-\la\,K_t^{\pm})\inv\,P^{\pm}+o_1(\la).\label{opest}\eq
Relations (\ref{resest}) with $\la=1$ follow from this since it may be
rewritten
\[P^{\pm}\,[(I-\la\,K_t)\inv-I]\,P^{\pm}=
P^{\pm}\,[(I-\la\,K_t^{\pm})\inv-I]\,P^{\pm}+o_1(\la),\]
and if we take the trace of both sides and divide by $\la$ we obtain
(\ref{resest}).\sp

Here is an outline of the proof of (\ref{opest}). We use the matrix
representations
of our operators corresponding to the decomposition of $L_2(\bR^+)$ as the
direct sum
of the spaces $L_2(0,1)$ and $L_2(1,\iy)$. Thus (the equal sign meaning ``has
matrix
representation'')
\[I-\la\,K_t=\ba I-\la\,P^-K_tP^- & -\la\,P^-K_tP^+\\&\\-\la\,P^+K_tP^- &
I-\la\,P^+K_tP^+\ea.\]
Because the nondiagonal corners of the matrix have the mutually orthogonal
projections
$P^{\pm}$ occurring as they do we will be able, with error $o_1(\la)$, to
replace the operator $K_t$
appearing there by the operator $K_0$ obtained from it by setting $t=0$. Thus
$K_0$ has
kernel
\bq K_0(u,v)=\uv\int_{\Om}{1\ov -\om u+v}\,d\rho(\om).\label{K0}\eq
(Note the lack of consistency with
the notation $K_k$ in the introduction; this should cause no confusion.)
If the diagonal entries $I-\la\,P^{\pm}K_tP^{\pm}$ are invertible, we can write
the resulting matrix as the product of
\[\ba I-\la\,P^-K_tP^- & 0\\&\\0 & I-\la\,P^+K_tP^+\ea\]
on the left and
\[\ba I & -\la\,(I-\la\,P^-K_tP^-)\inv
P^-K_0P^+\\&\\-\la\,(I-\la\,P^+K_t^+P^+)\inv P^+K_0P^- &
I\ea\]
on the right. Next, because of our assumption we shall be
able to show that the operators $I-\la\,P^{\pm}K_tP^{\pm}$ are uniformly
invertible for
small $t$ (i.e., the operator norms of their inverses are bounded) and that
their
inverses converge strongly to
$(I-\la\,P^{\pm}K_0P^{\pm})\inv$ as $t\ra0$. (Recall that $A_t$ is said to
converge strongly to $A$ as $t\ra0$ if $A_tf\ra Af$ for all $f$ in the
underlying space.)
This is actually the crux of the proof. After that it follows, because
$P^{\pm}K_0P^{\mp}$
are trace class, that with error $o_1(\la)$ we can replace
$(I-\la\,P^{\pm}K_tP^{\pm})\inv$
by $(I-\la\,P^{\pm}K_0P^{\pm})\inv$ in the nondiagonal entries. Thus, if we
define
\[\cM:=\ba I & -\la\,(I-\la\,P^-K_0P^-)\inv
P^-K_0P^+\\&\\-\la\,(I-\la\,P^+K_0^+P^+)\inv
P^+K_0P^- & I\ea,\]
we will have shown
\[I-\la\,K_t=\ba I-\la\,P^-K_tP^- & 0\\&\\0 &
I-\la\,P^+K_tP^+\ea\,\cM+o_1(\la).\]
 From this, using the uniform invertibility of the $I-\la\,P^{\pm}K_tP^{\pm}$
again and the
invertibility of the constant matrix $\cM$ (which we have to prove) we deduce
\[(I-\la\,K_t)\inv=\cM\inv\,\ba (I-\la\,P^-K_tP^-)\inv & 0\\&\\0 &
(I-\la\,P^+K_tP^+)\inv\ea+
o_1(\la).\]
Now (here is the trick), applying an analogous procedure to the operator family
$I-\la\,K_t^+$ gives
\[(I-\la\,K_t^+)\inv=\cM\inv\,\ba (I-\la\,P^-K_0P^-)\inv & 0\\&\\0 &
(I-\la\,P^+K_tP^+)\inv\ea+
o_1(\la).\]
It is clear that the lower-right entries of the two matrix products are
the same, and this is exactly the statement
\[ P^+\,(I-\la\,K_t)\inv\,P^+=P^+\,(I-\la\,K_t^+)\inv\,P^++o_1(\la),\]
which is half of (\ref{opest}). The other half is obtained similarly.\sp

Carrying out the details of the proof of this will require, first, some general
facts about
families of operators on a Hilbert space. \sp

\noi{\bf Fact 1}. If $A_t$ converges strongly to an invertible operator $A$
as $t\ra0$ and if the $A_t$ are uniformly invertible then $A_t\inv$ also
converges strongly
to $A\inv$.\sp

\noi This follows from the assumptions and the identity
$A_t\inv-A\inv=A_t\inv(A-A_t)A\inv$. The next fact says that strong convergence
can
sometimes be converted into trace norm convergence.\sp

\noi{\bf Fact 2}. If $A_t\ra A$ strongly and $B_t\ra B$ in trace norm then
$A_tB_t\ra AB$
in trace norm.\sp

\noi This is a variant of Proposition 2.1 of \cite{W1}. There the families of
operators
depended on a parameter $n\in{\bf Z}^+$ rather than $t\in\bR^+$, a matter of no
importance
since we may consider general sequences $t_n\ra0$. Also, instead of a sequence
of operators
converging in trace norm to $B$ there was the single trace class operator $B$.
The
apparently more general result follows trivially from this special case.\sp

\noi{\bf Fact 3}. Suppose $A_t$ and $A$ are as in Fact 1 and that $B_t$ are
trace class
operators converging in trace norm to $B$. Assume also that $A+B$ is
invertible. Then the
$A_t+B_t$ are uniformly invertible for sufficiently small $t$, and if
$B_t=o_1(1)$ then
$(A_t+B_t)\inv=A_t\inv+o_1(1)$.\sp

\noi{\bf Proof}. Write $A_t+B_t=A_t\,(I+A_t\inv B_t)$.
By Fact 1 $A_t\inv\ra A\inv$ strongly,
and so by Fact 2 (with $A_t$ repaced by $A_t\inv$) we deduce $I+A_t\inv B_t=
I+A\inv B+o_1(1)=(I+o_1(1))(I+A\inv B)$, since clearly
$(I+o_1(1))\inv=I+o_1(1)$. Both
statements now follow.\sp

In our derivation of (\ref{opest}) we have to know
that the $I-P^{\pm}K_t^{\pm}P^{\pm}$ are uniformly
invertible for small $t$. We shall deduce this from known facts about
uniform invertibility of truncated Wiener-Hopf operators, which we now
describe.
The proofs can be found in \cite{GF}.

The Wiener-Hopf operator $W$ associated with a function $k\in L_1(\bR)$ is the
operator
on $L_2(\bR^+)$ with kernel $k(x-y)$. Introduce the Fourier transform of $k$,
\[\hat{k}(\xi)=\int_{-\iy}^{\iy}e^{i\xi x}\,k(x)\,dx.\]
This is a continuous function on $\bR$ tending to 0 as $\xi\ra\pm\iy$. A
necessary
and sufficient condition that $I-W$ be invertible is that $1-\hat{k}(\xi)\neq0$
for all
$\xi$, and
\[\arg\,(1-\hat{k}(\xi))\Big|_{-\iy}^{\iy}=0.\]
The truncated Wiener-Hopf operators are the operators $P_{\al}WP_{\al}$ where
$P_{\al}$ denotes multiplication by the characteristic function of $(0,\al)$.
Clearly
these operators converge strongly to $W$ as $\al\ra\iy$. The important fact is
that if
$W$ is invertible, in other words if the conditions on $\hat{k}$ stated above
hold, then
the operators $I-P_{\al}WP_{\al}$ are uniformly invertible for sufficiently
large $\al$.
We mention also that the operator with kernel
$k(x-y)$ on the whole line $\bR$ is invertible if and only if the first
condtion
alone is satisfied, that $1-\hat{k}(\xi)\neq0$ for all $\xi$.
\par From this we can deduce information about kernels $k(u,v)$ which are
homogeneous of
degree $-1$ since the variable change $u=e^{x}$ transforms this kernel
into a convolution kernel. More precisely, denote by $U$ the unitary operator
from
$L_2(1,\iy)$ to $L_2(0,\iy)$ given by $Uf(x)=e^{x/2}f(e^x)$. Then if $T$
denotes
the operator on $L_2(1,\iy)$ with kernel $k(u,v)$, the operator $UTU\inv$ is
the
operator on $L_2(0,\iy)$ with kernel
\[e^{x/2}\,e^{y/2}\,k(e^x,e^y)=e^{(y-x)/2}\,k(1,e^{y-x}),\]
where we used the homogeneity of $k(u,v)$. Notice also that if $P_t^+$ denotes
multiplication by the characteristic function of $(1,t\inv)$ then $UP_t^+U\inv$
is the
projection operator $P_{\log t\inv}$ of the last paragraph. After making an
obvious
change of variable in computing the Fourier transform of
$e^{-x/2}\,k(1,e^{-x})$ we deduce
\sp

\noi{\bf Fact 4.} Assume that $\int_0^{\iy}v^{-1/2}\,|k(1,v)|\,dv<\iy$, denote
by $T$ the
operator with kernel $k(u,v)$ on $L_2(1,\iy)$ with kernel $k(u,v)$, and by
$M(s)$
the Mellin transform
\[M(s):=\int_0^{\iy}v^{s-1}\,k(1,v)\,dv.\]
Then a necessary and sufficient condition that $I-T$ be invertible is that
\bq 1-M(s)\neq0\ \mbox{for}\ \Re\,s=\ssl\hf,\ \ \
\arg\,(1-M(s))\Big|_{\hf-i\iy}^{\hf+i\iy}=0.
\label {Mconds}\eq
If this holds then the operators $I-P_t^+TP_t^+$ are uniformly invertible for
sufficiently
small $t$.\sp

\noi{\bf Remark.} If we use the variable change $u=e^{-x}$ instead of $u=e^x$
then our
operator $T$
acts on $L_2(0,1)$ and we find (again using homogeneity) that the condition for
invertibility of $I-T$ is exactly the same as before, and that if $P_t^-$
denotes
multiplication by the characteristic function of $(t,1)$ then this condition
implies the uniform invertibility of $I-P_t^-TP_t^-$ for sufficiently small
$t$.
Also (tranferring to this context the last sentence of the discussion of
Wiener-Hopf
operators), the same condition implies the invertibility of the operator $I-T$
on
$L_2(0,\iy)$.\sp

We apply this to the kernel $\la K_0$, which  is homogeneous of degree $-1$.
The relevant
Mellin transform is found to be
\[M(s)=\la\,{\pi\ov\sin\pi s}\,\int_{\Om}(-\om)^{s-\be-1}\,d\rho(\om).\]
so that $1-M(s)=h(s-\be)$, where $h$ is given by (\ref{h}). If we recall that
$\be=\hf-\sl$ we see that the conditions (\ref{Mconds}) are met, the first
immediately
from the
definition of $\sl$ and the second because the index (the variation of the
argument, which
is necessarily an integer) is a continuous function of $\la$, locally constant
and clearly
equal to 0 when $\la=0$.
So we know that the operators $I-\la\,K_0$ and $I-\la\,P^{\pm}K_0P^{\pm}$ are
invertible and
the operator families $I-\la\,P_t^{\pm}K_0P_t^{\pm}$ are uniformly invertible
for small $t$.\sp

We introduce one last piece
of notation. We denote by $K^{\pm}$ the operators on $L_2(\bR^+)$ with kernels
\[K^+(u,v)=\uv\int_{\Om}{e^{-(1-\om)\,u}\ov -\om u+v}\,d\rho(\om),\quad
K^-(u,v)=\uv\int_{\Om}{e^{-(1-\om\inv)\,u\inv}\ov -\om u+v}\,d\rho(\om).\]
Notice that rescaling $K^{\pm}(u,v)$
under the variable change $u\ra t^{\pm1}u$ gives $K_t^{\pm}(u,v)$.\sp

Next we derive various trace class properties
of our operators which will be needed. For these shall use an estimate for
the trace norm of an operator on $L_2(\bR^+)$ with kernel of the form
\bq \int_{\Om}{q_1(\om,u)\,q_2(\om,v)\ov -\om u+v}\,d\rho(\om)\label{qkern}\eq
which is a special case of the sublemma in
the appendix of \cite{W2}.
We denote by $\phi(s)$ any positive function on $\bR^+$, by $\Phi$ its
Laplace transform, and by $\Psi$ the Laplace transform of $\phi(s)\inv$. Then
there
is a constant $m$ depending on only on $\Om$ such that the trace norm of the
operator with
kernel given above is at most $m\inv$ times the square root of
\bq\int_0^{\iy}\int_{\Om}|q_1(\om,u)|^2\,\Phi(mu)\,d\rho(\om)\,du
\cdot\int_0^{\iy}\int_{\Om}\,|q_2(\om,u)|^2\,
\Psi(mu)\,d\rho(\om)\,du.\label{1est}\eq
 From this will follow our first lemma. In the proof we denote by $\ch^+$
the characteristic function of $(1,\iy)$ and by $\ch^-$
the characteristic function of $(0,1)$, so $P^{\pm}$ is multiplication
by $\ch^{\pm}$. \sp

\noi{\bf Lemma 1}. The operators (independent of $t$)
\[P^+K_0P^-,\ \ P^+K^+,\ \  P^-(K^+-K_0)\]
are trace class. The operators (depending on $t$)
\[P^+(K_t-K_0)P^-,\ \ P^-(K_t^+-K_0),\ \  P^+\,(K_t-K_t^+)\]
are $o_1(1)$ as $t\ra0$. The statements also hold if all superscripts $+$ and
$-$ are
interchanged.\sp

\noi{\bf Proof}. All the operators have kernel of the form (\ref{qkern}). We
list the
operators below, together with the corresponding functions $q_1(\om,u)$ and
$q_2(\om,u)$.
\[\begin{array}{lll}
\mbox{Operator} & q_1(\om,u) & q_2(\om,u)\\&&\\
P^+K_0P^- & \ch_{(1,\iy)}(u)\ub & \ch_{(0,1)}(u)\um\\
P^+K^+ & \ch_{(1,\iy)}(u)\,e^{-(1-\om)\,u}\ub & \um\\
P^-(K^+-K_0) & \ch_{(0,1)}(u)\,(e^{-(1-\om)\,u}-1)\ub & \um\\
P^+(K_t-K_0)P^- &
\ch_{(1,\iy)}(u)\,(e^{-t\,[(1-\om)\,u+(1-\om\inv)\,u\inv]}-1)\ub &
\ch_{(0,1)}(u)\um\\
P^-(K_t^+-K_0) & \ch_{(0,1)}(u)\,(e^{-t(1-\om)\,u}-1)\ub & \um\\
P^+\,(K_t-K_t^+) &
\ch_{(1,\iy)}(u)\,e^{-t(1-\om)\,u}\,(e^{-t(1-\om\inv)\,u\inv}-1)\ub & \um
\end{array}\]
For each of these operators we take two numbers $p,q\in(-1,1)$ and define
$\phi(s)=s^p$ for $s\leq 1$ and $\phi(s)=s^q$ for $s\geq 1$. We easily see that
\[\Phi(s)=\left\{\begin{array}{ll}
O(u^{-p-1}) & {\mbox for}\ u\geq1\\&\\O(u^{-q-1}) & {\mbox for}\
u\leq1,\end{array}\right.
\ \ \Psi(s)=\left\{\begin{array}{ll}
O(u^{p-1}) & {\mbox for}\ u\geq1\\&\\O(u^{q-1}) & {\mbox for}\
u\leq1.\end{array}\right.\]
For each operator one can find $p$ and $q$ such that both integrals in
(\ref{1est})
are finite, and any integral depending on $t$ is $o(1)$ as $t\ra0$. In fact, as
the
reader can check, we may take for all the operators any $q\in(2\be,2\be+2)$,
for the
first and fourth operators any $p>2\be$ and for the other four any
$p\in(2\be-2,2\be)$.
This takes care of the six displayed operators. For the other
six we use the fact that the substitutions $u\ra u\inv,\ \om\ra\om\inv$ yield
operators
of the same form with $\be$ replaced with $-\be$ and with the superscripts
interchanged.\sp

As a preliminary to the next lemma we show that certain modified Laplace
tranforms
are bounded operators on $L_2(\bR^+)$.\sp

\noi{\bf Lemma 2}. The integral operator on $L_2(\bR^+)$ with kernel
$(uv)^{-\be}\,e^{-uv}$ is bounded if $\be<\hf$.\sp

\noi{\bf Proof}. The mapping $f(v)\ra v\inv f(v\inv)$ is unitary, so we may
replace the kernel by $u^{-\be}v^{\be-1}\,e^{-u/v}$. Under the unitary mapping
$f(u)\ra e^{x/2}\,f(e^x)$ this becomes the kernel
$e^{(\hf-\be)(x-y)}e^{-e^{x-y}}$
on $(-\iy,\iy)$.
Thus the operator becomes convolution by the $L_1$ function
$k(x)=e^{(\hf-\be)x}e^{-e^x}$ and so is bounded.\sp

\noi{\bf Lemma 3}. The operators $I-\la\,K^{\pm}$ are invertible.\sp

\noi{\bf Proof}. We consider $K^+$, which we can write as $P^-K_0+P^-(K^+-K_0)
+P^+K^+$. By the first part of Lemma 1 the second summands are trace class,
therefore
certainly compact. Our assumption implies that $I-\la\,P^-K_0$ is invertible.
Hence
$I-\la\,K^+$ is the sum of an invertible operator and a compact operator, and
so it follows
from general theory that it will be invertible if 0 is not an eigenvalue. In
other
words it suffices to prove that $\la\,K^+f=f$ for $f\in L_2(\bR^+)$ implies
$f=0$.

For any $\om\in \bC\backslash\bR^+$ and any $f\in L_2(\bR^+)$ we have for $x>0$
\[\int_0^{\iy}e^{-xu}u^{-\be}\,du\int_0^{\iy}\uv\,{e^{-(1-\om)\,u}\ov -\om
u+v}\,f(v)\,dv\]
\[=\int_0^{\iy}{dy\ov x+1-\om(y+1)}\,\int_0^{\iy}e^{-yv}\,v^{-\be}f(v)\,dv.\]
This can be seen for $\Re\,\om<0$ by using the integral reresentation
\[{1\ov -\om u+v}=\int_0^{\iy}\,e^{-y(-\om u+v)}\,dy\]
in the integral on the left above and interchanging the order of integration.
The
identity follows for all $\om\in\Om$ since both sides are analytic functions
of $\om$ in this domain. Now suppose that $\la\,K^+f=f$. Then if we
integrate both sides of the identity with respect to $d\rho(\om)$ and
multiply by $\la$ the left side becomes the Laplace transform of
$u^{-\be}\,f(u)$, which we
denote by $g(x)$, and the
the right side becomes
\[\la\int_0^{\iy}\int_{\Om}{d\rho(\om)\ov x+1-\om(y+1)}\,g(y)\,dy.\]
Using the fact $y^{-\be}g(y)\in L_2$, which we know by the previous lemma, we
see that
the integral is a bounded function of $x$. If we recall the definition
(\ref{K0}) then we see that the identity becomes
\[g(x)=\la\,\int_0^{\iy}\Big({x+1\ov y+1}\Big)^{\be}K_0(y+1,x+1)\,g(y)\,dy\ \ \
\,(x\geq0),\]
or
\[x^{-\be}\,g(x-1)=\la\,\int_1^{\iy}K_0(y,x)\,g(y-1)\,y^{-\be}\,dy\ \ \ \
(x\geq1).\]
Now we know that $x^{-\be}\,g(x)$ is in $L_2(0,\iy)$ and that $g(x)$ is
bounded. It follows
that $x^{-\be}\,g(x-1)$ belongs to $L_2(1,\iy)$. The right side above is the
operator
with kernel $K_0(y,x)$ acting on this function. Thus (if $f\neq0$) the operator
$\la\,P^+K_0'P^+$ has 1 as an eigenvalue, where $'$ denotes transpose, so
$I-\la\,P^+K_0'P^+$ is not invertible. But this implies
$I-\la\,P^+K_0P^+$ is not invertible, whereas we know that it is. This
contradiction establishes the lemma.\sp

\noi{\bf Lemma 4}. The operators $I-\la\,P^{\pm}K_t^{\pm}P^{\pm}$ are uniformly
invertible
for sufficiently small~$t$.\sp

\noi{\bf Proof}. We consider $I-\la\,P^+K_t^+P^+$ and for this it is enough to
show that the
$I-\la P^+K_t^+$ are uniformly invertible. The kernel of $P^+K_t^+P^+$ is
$\ch_{(1,\iy)}(u)\,K_t^+(u,v)$ and the
substitution $u\ra t\inv u$ allows us to consider instead the operator with
kernel
$\ch_{(t,\iy)}(u)\,K^+(u,v)$. We write this (not displaying the variables $u$
and
$v$) as
\[\ch_{(t,1)}\,K_0+\ch_{(1,\iy)}\,
K^++\ch_{(0,1)}\,(K_+-K_0)+\ch_{(0,t)}\,(K_0-K^+).\]
Recalling the definitions of our various projection operators we see that the
first
kernel corresponds to the operator $P_t^-K_0$, and we know that the
$I-\la\,P_t^-K_0$ are uniformly invertible for sufficiently small $t$.
The second and third summands correspond to the operators $P^+K^+$ and
$P^+(K^+-K_0)$, which we know by Lemma 1 to be trace class. The last summand
corresponds
to the operator $P^-(K_0-K^+)$, which we know by Lemma 1 to be trace class,
left-multiplied by multiplication by $\ch_{(0,t)}$, which converges strongly to
0.
An application of Fact 2 shows that this last operator is $o_1(1)$. The strong
limit of the
sum of the four operators is, of course, $K^+$ and we know by Lamma 2 that
$I-\la\,K^+$
is invertible. Hence we can apply Fact 3 to deduce the the result. \sp

We can now fill in the details of the proof of (\ref{opest}) outlined earlier.
Thus
we begin with the representation
\[I-\la\,K_t=\ba I-\la\,P^-K_tP^- & -\la\,P^-K_tP^+\\&\\-\la\,P^+K_tP^- &
I-\la\,P^+K_tP^+\ea.\]
Applying Lemma 1 to the nondiagonal entries we deduce
\[I-\la\,K_t=\ba I-\la\,P^-K_t^-P^- & -\la\,P^-K_0P^+\\&\\-\la\,P^+K_0P^- &
I-\la\,P^+K_t^+P^+\ea+o_1(\la).\]
Lemma 3 tells us in particular that the diagonal entries of this matrix are
invertible for small $t$ so we may factor out
\[\ba I-\la\,P^-K_tP^- & 0\\&\\0 & I-\la\,P^+K_tP^+\ea\]
on the left, leaving
\[\ba I & -\la\,(I-\la\,P^-K_t^-P^-)\inv P^-K_0P^+\\&\\
-\la\,(I-\la\,P^+K_t^+P^+)\inv P^+K_0P^- & I\ea\]
on the right. Next we combine the uniform invertibility of the
$I-\la\,P^{\pm}K_t^{\pm}P^{\pm}$ proved in
Lemma 3 with Fact 1 to deduce that the inverses of these operators converge
strongly to
$(I-\la\,P^{\pm}K_0^{\pm}P^{\pm})\inv$. Since $P^{\pm}K_0P^{\mp}$ are trace
class, by
Lemma 1, we deduce by Fact~2 that the matrix above is $\cM+o_1(\la)$ where
$\cM$ is the
matrix obtained by replacing $K_t^{\pm}$ by $K_0$. Thus
\[I-\la\,K_t=\ba I-\la\,P^-K_tP^- & 0\\&\\0 &
I-\la\,P^+K_tP^+\ea\,\cM+o_1(\la).\]

Now we have to know that $\cM$ is invertible, and we see
this as follows. If, instead of the operator $I-\la\,K_t$ which depends on $t$,
we had
started with the operator $I-K_0$ then we would have obtained the exact
representation
\[I-\la\,K_0=\ba I-\la\,P^-K_0P^- & 0\\&\\0 & I-\la\,P^+K_0P^+\ea\,\cM.\]
Since both $I-\la\,K_0$ and the matrix on the left are invertible, by our
assumption, we
deduce that $\cM$ is invertible.

Next we go through a similar process starting with the
operator $I-\la\,K_t^+$ rather than  $I-\la\,K$. Using the fact that
$P^-K_t^+=P^-K_0+o_1(1)$, which we know by Lemma~1, we obtain in this case
\[I-\la\,K_t^+=\ba I-\la\,P^-K_0P^- & 0\\&\\0 &
I-\la\,P^+K_tP^+\ea\,\cM+o_1(\la).\]
 From these matrix representations and the facts that $\cM$ and
$I-\la\,P^-K_0P^-$ are
invertible and $I-P^{\pm}K_t^{\pm}P^{\pm}$ uniformly invertible we deduce,
using Fact 3 with $o_1(1)$ replaced by $o_1(\la)$,
\[(I-\la\,K_t)\inv=\cM\inv\,\ba (I-\la\,P^-K_tP^-)\inv & 0\\&\\0 &
(I-\la\,P^+K_tP^+)\inv\ea+o_1(\la),\]
\[(I-\la\,K_t^+)\inv=\cM\inv\,\ba (I-\la\,P^-K_0P^-)\inv & 0\\&\\0 &
(I-\la\,P^+K_tP^+)\inv\ea+o_1(\la).\]
Comparing lower-right entries of the matrices gives
\[ P^+\,(I-K)\inv\,P^+=P^+\,(I-K_t^+)\inv\,P^++o_1(\la),\]
which is half of (\ref{opest}). The other half is obtained similarly.\sp

\noi{\bf Remark}. To apply (\ref{resest}) to (\ref{detrep}) we need something
extra,
e.g., that (\ref{resest}) holds uniformly for these $\la$. With a little care
our argument
gives this also, but we spare the reader the details.\sp

\setcounter{equation}{0}\renewcommand{\theequation}{3.\arabic{equation}}

\noi{\bf 3. The resolvents of}  {\boldmath$K^{\pm}$}\sp

We are going to find integral representations for the integrals on the right
side of
(\ref{resest}), and we consider $\int_1^{\iy}R_t^+(u,u;\la)\,du$ first. The
substitution
$u\ra u/t$ shows that this equals $\int_t^{\iy}R^+(u,u;\la)\,du$ where
$R^+(u,v;\la)$ is the
resolvent kernel of the operator $K^+$. For this we require only that $\Om$ be
a compact
subset of
\bq\{\om\in\bC:\,\Re\,\om<1,\ \om\not\in \bR^+\},\label{Om}\eq
since the term $1-\om\inv$ does not appear in the exponent in the kernel of
$K^+$.
The derivation will involve an initial step
which is valid only when $\Om$ is contained in the left half-plane so we
assume this to begin with. We shall also assume that $\la$ is so small that
$h(s)\neq$ for $\Re\,s=\hf$, so that with the notation of the last section we
may
take $\sl=\hf,\ \be=0$. Eventually these two assumptions will be removed by an
analytic
continuation argument.

Because $\be=0$ the kernel of $K^+$ is
\bq K^+(u,v)=\int_{\Om}{e^{-t(1-\om)u}\over -\om u+v}\,d\rho(\om).\label{G}\eq
If we set
\[A(u,x):=\int_{\Om}\,e^{-(1-\om)u}\,e^{\om ux}\,d\rho(\om),\quad
B(x,u):=e^{-ux},\]
then
\[K^+(u,v)=\int_0^{\iy}\,A(u,x)\,B(x,v)\,dx.\]
Lemma 2 of the preceding section tells us that $B(u,x)$ is the kernel of a
bounded operator
from $L_2(\bR^+)$ to $L_2(\bR^+)$ and, with our assumption on $\Om$, that the
same is
true of $A(x,u)$. The above shows that $K^+=AB$, and the operator $BA$ has
kernel
\[\int_0^{\iy}B(x,u)\,A(u,y)\,du=\int_{\Om}{d\rho(\om)\ov x-\om y+1-\om}
=\int_{\Om}{d\rho(\om)\ov (x+1)-\om(y+1)}.\]
We use the general fact $AB(I-\la AB)\inv=A(I-\la BA)\inv B$ to deduce that
$R^+(u,u)$ is
given by an inner product,
\bq R^+(u,u)=\Big((I-\la BA)\inv
B(\,\cdot\,,u),A(u,\,\cdot\,)\Big).\label{Ru}\eq
We begin by computing
\[f:=(I-\la BA)\inv B(\,\cdot\,,u).\]
Thus we want to solve
\[f(x)-\la\,\int_0^{\iy}\int_{\Om}{d\rho(\om)\ov
(x+1)-\om(y+1)}\,f(y)\,dy=e^{-ux}
\qquad (x\geq0),\]
or
\[f(x-1)-\la\,\int_1^{\iy}\int_{\Om}{d\rho(\om)\ov x-\om
y}\,f(y-1)\,dy=e^{-u(x-1)}\qquad
(x\geq1).\]
The substition $x\ra e^x$ brings this to the form of a Wiener-Hopf equation,
so we can use the factorization method to find the solution.

We begin by decreeing that the last indentity holds for all $x\geq0$, in other
words we
define $f$ on $(-1,0)$ by the identity. Then we define
\[F_-(s):=\int_1^{\iy}x^{s-1}\,f(x-1)\,dx,\qquad
F_+(s):=\int_0^1x^{s-1}\,f(x-1)\,dx.\]
These belong to the spaces Hardy spaces $H_2(\Re\,s<\hf),\ H_2(\Re\,s>\hf)$,
respectively.
We take Mellin transforms of both sides of the equation, and find that for
$\Re\,s=\hf$
\[F_-(s)+F_+(s)-{\la\,\pi\ov \sin\,\pi
s}\int_{\Om}(-\om)^{s-1}\,d\rho(\om)\,F_-(s)=
e^u\,u^{-s}\,\Ga(s).\]
(The exponential in the integral is made definite by taking
$|\arg(-\om)|<\pi$.) We
write this as
\bq H(s)\,F_-(s)+F_+(s)=e^u\,u^{-s}\,\Ga(s),\label{rh}\eq
where
\bq H(s):={h(s)\ov \sin\,\pi s}=1-{\la\,\pi\ov \sin\,\pi
s}\int_{\Om}(-\om)^{s-1}\,
d\rho(\om).\label{H}\eq
This function is bounded and analytic in each vertical strip of the complex
$s$-plane,
away from the zeros of $\sin\,\pi s$, $H(s)-1\ra 0$ exponentially as $\Im
s\ra\pm\iy$ and
\[\arg\,H(s)\Big|_{\hf-i\iy}^{\hf+i\iy}=0.\]
Thus there is be a representation
\[H(s)={H_-(s)\over H_+(s)},\]
where $H_-(s)^{\pm1}$ are bounded and analytic in $\Re\,s\leq\hf+\dl$ for some
$\dl>0$ and
$H_+(s)^{\pm1}$ are bounded and analytic in $\Re\,s\geq\hf-\dl$. We multiply
(\ref{rh})
by $H_+(s)$ and use the decomposition $F=F_-+F_+$ of an arbitrary function in
$L_2(\Re\,s=\hf)$ into boundary functions of functions in $H_2(\Re\,s<\hf)$ and
$H_2(\Re\,s>\hf)$ to write the result as
\[H_-(s)\,F_-(s)-e^u\,\Big(u^{-s}\,\Ga(s)\,H_+(s)\Big)_-=
-H_+(s)\,F_+(s)+e^u\,\Big(u^{-s}\,\Ga(s)\,H_+(s)\Big)_+.\]
The two sides are boundary functions of functions in $H_2(\Re\,s<\hf)$ and
$H_2(\Re\,s>\hf)$, respectively, so they both vanish. This gives the
representation
\bq F_-(s)={e^u\over H_-(s)}\,\Big(u^{-s}\,\Ga(s)\,H_+(s)\Big)_-.\label{F}\eq

Now (see (\ref{Ru})) we have to multiply $f(x)$ by $A(u,x)$ and integrate with
respect
to $x$ over $(0,\iy)$. This is
\[\int_{\Om}d\rho(\om)\int_0^{\iy}f(x)\,e^{-(1-\om)u}\,e^{\om ux}\,dx=
e^{-u}\,\int_{\Om}d\rho(\om)\int_0^{\iy}f(x-1)\,\ch_{(1,\iy)}(x)\,e^{\om
ux}\,dx.\]
The Mellin transform of
$f(x-1)\,\chi_{(1,\iy)}(x)$ equals $F_-(s)$ and the Mellin transform of $e^{\om
ux}$
equals $(-\om u)^{-s}\,\Ga(s)$, so Parseval's formula for Mellin transforms
shows
that the above equals
\[e^{-u}\,\int\int_{\Om}(-\om
u)^{s-1}\,d\rho(\om)\,\Ga(1-s)\,F_-(s){ds\over2\pi i},\]
the outer integration taken over $\Re\,s=\hf$. (All vertical integrals are
taken in the direction
from $-i\iy$ to $i\iy$.)
Next we recall (\ref{F}) and use the integral representation of the operator
$G\ra G_-$ to
write the above as
\[\int\int_{\Om}(-\om u)^{s-1}\,d\rho(\om)\,{\Ga(1-s)\over
H_-(s)}\,{ds\over2\pi i}\int
{u^{-s'}\,\Ga(s')H_+(s')\over s'-s}\,{ds'\over2\pi i},\]
the inner integral taken over $\Re\,s'=\hf+\dl$. Alternatively, this may be
written
\[\int\int_{\Om}(-\om)^{s-1}\,d\rho(\om)\,{\Ga(1-s)\over H_-(s)}\,{ds\over2\pi
i}\int
{u^{-s'-1}\,\Ga(s'+s)H_+(s'+s)\over s'}\,{ds'\over2\pi i},\]
where now the inner integral is taken over $\Re\,s'=\dl$.
The integrands of these integrals
vanish exponentially at infinity, and $u$ occurs to the power $-s'-1$, which
has real part
$-\dl-1$. Thus we may integrate with respect to $u$ from $t$ to $\iy$ under the
integral signs and deduce that
\[\int_t^{\iy}R^+(u,u;\la)\,du=
\int\int_{\Om}(-\om)^{s-1}\,d\rho(\om)\,{\Ga(1-s)\over H_-(s)}\,
{ds\over2\pi i}\int{t^{-s'}\,\Ga(s'+s)H_+(s'+s)\over s'^2}\,{ds'\over2\pi i}.\]
It follows from (\ref{H}), and the gamma function representation of the last
factor
there, that
\[\int_{\Om}(-\om)^{s-1}\,d\rho(\om)\,\Ga(1-s)=\la\inv\,(1-H(s))\,\Ga(s)\inv.\]
Thus we have shown (reverting to the resolvent $R_t^+$) that
\[\la\,\int_1^{\iy}R_t^+(u,u;\la)\,du\]
\[=\int_{\hf-i\iy}^{\hf+i\iy}(H_-(s)\inv-H_+(s)\inv)\,\Ga(s)\inv\,
{ds\over2\pi i}\int_{\dl-i\iy}^{\dl+i\iy}{t^{-s'}\,\Ga(s'+s)H_+(s'+s)\over
s'^2}\,
{ds'\over2\pi i}.\]

This was proved if $\la$ is sufficiently small and if $\Om$ lies in the left
half-plane.
Let us remove the latter condition first. For any $\eta$ define
the measure $\rho_{\eta}$ by $\rho_{\eta}(E)=\rho(E-\eta)$. This has support
$\Om+\eta$.
For all $\eta$ in a neighborhood in $\bC$ of $[-1,\,0]$ the set $\Om+\eta$ is
contained in
the region (\ref{Om}).
For $\eta$ near $-1$ the set will also lie in the left half-plane.
If $\la$ is small enough the condition on the zeros will be satisfied for the
measures $\rho_{\eta}$ for all $\eta$ in a neighborhood of $[-1,\,0]$.
For such $\la$ we know that the above formula holds for
$\eta$ in a neighborhood of $-1$. But both sides
are analytic functions of $\eta$ in our neighborhood. Thus the
formula must hold for $\eta=0$ also, which is what we wanted to show.

To remove the condition that $\la$ be small we must modify the formula to read
\[\la\,\int_1^{\iy}R_t^+(u,u;\la)\,du\]
\bq =\int_{\sl-i\iy}^{\sl+i\iy}(H_-(s)\inv-H_+(s)\inv)\,\Ga(s)\inv\,
{ds\over2\pi i}\int_{\dl-i\iy}^{\dl+i\iy}{t^{-s'}\,\Ga(s'+s)H_+(s'+s)\over
s'^2}\,
{ds'\over2\pi i},\label{Rint}\eq
where $\sl\in (0,1)$ is as in the previous section, a continuously varying
function
of $\la$ such that $1-H(s)$ is nonzero on the line $\Re\,s=\sl$. Both sides of
(\ref{Rint})
are analytic functions of $\la$ for $\la$ in a neighborhood of our path, they
agree
near $\la=0$, so they agree everywhere on the path.

As for $R^-(u,u;\la)$ on $(0,1)$ the change of variable $u\ra u\inv$
transforms the kernel $K^-(u,v)$ into
\[\tl K^+(u,v):=\int_{\Om}{e^{-t(1-\om\inv)u}\over -\om v+u}\,d\rho(\om)
=\int_{\Om}{e^{-t(1-\om)u}\over -\om u+v}\,(-\om)\,d\rho(\om\inv).\]
Therefore
\[R^-(u,u;\la)=u^{-2} \tl R^+(u\inv,u\inv;\la),\]
where $\tl R^+$ is the resolvent kernel for $\tl K^+$, and so
\[\int_0^1R^-(u,u;\la)\,du=\int_1^{\iy}\tl R^+(u,u;\la)\,du.\]
It is easy to see that replacing $K^+$ by $\tl K^+$ replaces $H(s)$ by $H(1-s)$
and so the integral is equal to
(\ref{Rint}) with $H(s)$ replaced by $H(1-s)$ and $\sl$ replaced by $1-\sl$.\sp

Now that we have these explicit representations it is obvious what we do: in
the inner
integral in (\ref{Rint}) we move the line of
$s'$-integration from $\Re\,s'=\dl$ to $\Re\,s'=-\dl$. We can do this if $\dl$
is small
enough. The residue at the double pole at $s'=0$ contributes
\bq-\int\,(H(s)\inv-1)\,{ds\over2\pi i}\,\log\,t
+\int\,(H_-(s)\inv-H_+(s)\inv)\,{1\over\Ga(s)}(\Ga(s)\,H_+(s))\,'\,{ds\over2\pi
i},
\label{intform}\eq
the integrations taken over $\Re\,s=\sl$,
and the error term is $O(t^{\dl})$. We do the same with $H(s)$ replaced by
$H(1-s)$
and add, and so we have obtained the asymptotics of
$\la\int_0^{\iy}R(u,u;\la)\,du$.\sp

\setcounter{equation}{0}\renewcommand{\theequation}{4.\arabic{equation}}

\noi{\bf 4. Asymptotics of det\,{\boldmath$(I-K)$}\,---The periodic case}\sp

The formula for $H(s)$ is now
\[ H(s)=1-{\la\,\pi\ov \sin\,\pi s}\sum_{\Om}c_{\om}\,(-\om)^{s-1},\]
$\om$ running over the $n$th roots of unity other than 1.
If in our sums we set $-\om=e^{{\pi i\over n}(2j-n)},\ (j=1,\cdots,n-1)$ then
$|\arg\,(-\om)|=|{\pi\over n}(2j-n)|<\pi$ as required. If we also
set $z=e^{{\pi i\over n}s}$ then the above may be written as
\[\sin\pi s\,H(s)=\sin\,\pi s+\la\pi\sum_{j=1}^{n-1}\,\om\inv
c_{\om}\,z^{2j-n},\]
or
\[2i\,\sin\pi s\,H(s)=z^{-n}\,\Big[z^{2n}-1+2i\la\pi\sum_{j=1}^{n-1}\,\om\inv
c_{\om}
\,z^{2j}\Big].\]
Recall that $\sin\pi s\,H(s)=h(s)$. The expression in brackets above is a
polynomial of
degree $n$ in $z^2$ and its zeros are the quantities
by $e^{{2\pi i\over n}\al}$, where $\al$ runs through the zeros
$\al_k=\al_k(\la)\ (k=
1,\cdots,n)$. With this notation the right side above is equal to
\bq z^{-n}\,\prod_{\al}(z^2-e^{{2\pi i\over n}\al})=
\prod_{\al}(z\,e^{-{\pi i\over n}\al}-z\inv\,e^{{\pi i\over n}\al})\,
\prod_{\al}\,e^{{\pi i\over n}\al}.\label{Hrep}\eq
Here and below the index $\al$ runs over the set $\{\al_1,\cdots,\al_n\}$.
The last product, a square root of the product of the roots of the polynomial,
equals
$\pm1$ or $\pm i$. This product equals $e^{{\pi i\over n}(1+\cdots+n)}=i^{n+1}$
for
$\la=0$ and so for all $\la$. Recalling that
$z=e^{{\pi i\over n}s}$ we see that we have obtained the representation
\[H(s)={(-1)^n\,2^{n-1}\over\sin\pi s}\,\prod_{\al}\,\sin{\pi\over n}(s-\al).\]

We now evaluate the integrals in (\ref{intform}). For $\la$ sufficiently small
again, the $\al_k$ will all lie in the strip $\hf<\Re\,s<n+\hf$. We may assume
this since
the usual analytic continuation will give the general case.

To evaluate the first integral in (\ref{intform}) we consider
\[\int\,(H(s)\inv-1)\,s\,{ds\over2\pi i}\]
taken over the infinite rectangle which is the contour running from
$n+\hf-i\iy$ to
$n+\hf+i\iy$
and then from $\hf+i\iy$ to $\hf-i\iy$. On the one hand this equals $n$ times
the first
integral in (\ref{intform}), and on the other hand it equals the sum of the
residues at
the poles between the two lines. Thus we have shown that
\bq\int\,(H(s)\inv-1)\,{ds\over2\pi i}={1\over
n}\,\sum_{\al}\,\al\,H'(\al)\inv.\label{aint}\eq

For the second integral in (\ref{intform}) we have to write down the explicit
expression
for the factors $H_{\pm}(s)$. These are given by
\[H_+(s)={\prod_\al\,\Ga({s-\al\over n}+1)\over\Ga(s)}\,n^s,\ \ \
H_-(s)={(-1)^n\,2^{n-1}\,\Ga(1-s)\over\prod_\al\,\Ga(-{s-\al\over n})}\,n^s.\]
It is readily verified that $H(s)=H_-(s)/H_+(s)$ and
that $H_-(s)^{\pm1}$ and $H_+(s)^{\pm1}$ are are bounded and analytic in
$\Re\,s\leq\hf+\dl$ and $\Re\,s\geq\hf-\dl$, respectively, for small $\la$.
Thus, they are the correct factors. The second integral in (\ref{intform}) may
be written
\[\int\,(H(s)\inv-1)\,{(\Ga(s)\,H_+(s))'\over\Ga(s)\,H_+(s)}\,{ds\over2\pi
i},\]
and by above expression for $H_+(s)$ this equals
\[\int\,(H(s)\inv-1)\,\Big[{1\over n}\sum_{\al'}{\Ga'({s-\al'\over n}+1)\over
\Ga({s-\al'\over n}+1)}+\log\,n\Big]\,{ds\over2\pi i},\]
where in the sum $\al'$ also runs over the set $\{\al_1,\cdots,\al_n\}$.
The contribution of the term $\log n$ is exactly $\log n$ times (\ref{aint}).
We evaluate the rest of this integral we
use the characteristic property of the Barnes $G$-function,
$G(z+1)=\Ga(z)\,G(z)$. Putting
$z$ equal to ${s-\al'\over n}+1$ and taking logarithmic derivatives gives
\[{G'({s-\al'\over n}+2)\over G({s-\al'\over n}+2)}-{G'({s-\al'\over n}+1)
\over G({s-\al'\over n}+1)}=
{\Ga'({s-\al'\over n}+1)\over \Ga({s-\al'\over n}+1)}.\]
We integrate
\[(H(s)\inv-1)\,\sum_{\al'}\,{G'({s-\al'\over n}+1)\over G({s-\al'\over
n}+1)}\]
over the same infinite rectangle as before. (This is justified by the fact that
$H(s)\inv-1$ vanishes exponentially at $\iy$ in vertical strips while
$G'(z)/G(z)$ grows like $z\log\,z$.) By the above relation the result is
exactly the
integral we want, and so computing residues gives the formula
\[{1\over n}\int\,(H(s)\inv-1)\,\sum_{\al'}{\Ga'({s-\al'\over n}+1)\over
\Ga({s-\al'\over n}+1)}\,{ds\over2\pi i}=
{1\over n}\sum_{\al,\al'}{G'({\al-\al'\over n}+1)\over G({\al-\al'\over
n}+1)}\,
H'(\al)\inv.\]

Thus, we have shown that
\[\la\,\int_1^{\iy}R_t^+(u,u,\la)=a^+(\la)\,\log\Big({t\ov
n}\Big)+b^+(\la)+O(t^{\dl}),\]
where
\bq a^+(\la)=-{1\over n}\,\sum_{\al}\,\al\,H'(\al)\inv,\ \ b^+(\la)={1\over
n}\sum_{\al,\al'}
{G'({\al-\al'\over n}+1)\over G({\al-\al'\over n}+1)}\,
H'(\al)\inv.\label{abp}\eq
We must add to this $\la\,\int_0^1R_t^-(u,u,\la)\,du$
which, as was mentioned earlier, is obtained by replacing $H(s)$ by $H(1-s)$.
The zeros
of this function which lie near $1,\cdots, n$ for small $\la$ are $n-\al+1$.
Hence
(\ref{abp}) is replaced by
\[a^-(\la)={1\over n}\,\sum_{\al}\,(n-\al+1)\,H'(\al)\inv,\]
\[b^-(\la)=-{1\over n}\sum_{\al,\al'}{G'({\al'-\al\over n}+1)\over
G({\al'-\al\over n}+1)}\,
H'(\al)\inv=
-{1\over n}\sum_{\al,\al'}{G'({\al-\al'\over n}+1)\over G({\al-\al'\over
n}+1)}\,
H'(\al')\inv.\]
Here we have used the periodicity of $H$ and the fact
${d\over ds}H(1-s)=-H'(1-s)$.
Adding and using (\ref{resest}), we see that
\bq\la\,\int_0^{\iy}R(u,u;\la)\,du=a(\la)\,\log\Big({t\ov
n}\Big)+b(\la)+O(t^{\dl}).\label{Rasym}\eq
where $a(\la)=a^+(\la)+a^-(\la),\ b(\la)=b^+(\la)+b^-(\la)$.

Now to obtain the asymptotics of $\log\,\det\,(I-K)$ we must replace $\la$ by
$\mu$,
multiply the above by
$-d\mu/\mu$ and integrate from 0 to $\la$. (Notice the factor $\la$ on the left
side
of (\ref{Rasym}) and recall
the minus sign in (\ref{detrep}).) We obtain from (\ref{H}) that for a zero
$\al(\la)$ of $H$ we have $\la\,d\al/d\la=H'(\al)\inv$.
Thus the coefficient $a(\la)$ in (\ref{Rasym}) may be written (after replacing
$\la$ by
$\mu$ and thinking of $\al$ as $\al(\mu)$)
\[-{1\over n}\sum_{\al}\,(2\al-n-1)\,\mu\,d\al/d\mu.\]
Multiplying by $-d\mu/\mu$ and integrating gives
\[{1\over n}\sum_{\al}\,(\al^2-(n+1)\,\al)\,\Big|_{\mu=0}^{\mu=\la}=
{1\over n}\sum_{\al}\,\al^2\,\Big|_{\mu=0}^{\mu=\la}\]
since, as we have already seen, $\sum\al$ is independent of $\la$.
Similarly $b(\mu)$ may be written
\[{1\over n}\sum_{\al,\al'}{G'({\al-\al'\over n}+1)\over G({\al-\al'\over
n}+1)}
\mu\,d(\al-\al')/d\mu\]
and multiplying by $-d\mu/\mu$ and integrating gives
\[-\sum_{\al,\al'}\log\,G({\al-\al'\over n}+1)\,\Big|_{\mu=0}^{\mu=\la}.\]
If we recall that when $\mu=0$ the zeros are $1,\cdots, n$
we see that the formulas for the constants in (\ref{detform}) are
the ones stated in the introduction.\sp

To obtain the asymptotics of the $q_k(t)$ we must consider $\det\,(I-\la\,K_k)$
instead of $\det\,(I-\la\,K)$. This amounts to replacing the coeffiients
$c_{\om}$ by
$\om^k\,c_{\om}$, and this in turn amounts to replacing $H(s)$ by $H(s+k)$. The
zeros
of this function modulo $n$ are $\al_1(\la)-k,\cdots, \al_n(\la)-k$. But these
are not the
zeros which are
to replace the $\al,\ \al'$ in our formulas for $a$ and $b$ since they do not
arise from
from the zeros whose values are $1,\cdots,n$ when $\la=0$. Rather, the
replacements must be
\[\al_{k+1}(\la)-k,\cdots,\al_n(\la)-k,\
\al_1+n-k(\la),\cdots,\al_k+n-k(\la),\]
which are the zeros with this property. Thus, for the asymptotics of $q_k(t)$
we make
these replacements in our formulas and the corresponding replacements with
$k-1$ instead of
$k$, subtract, and take logarithms. The result is found, after some computation
and the use of the functional equation for the $G$-function, to be the
asymptotics stated
in the introduction.\sp

\setcounter{equation}{0}\renewcommand{\theequation}{5.\arabic{equation}}

\noi{\bf 5. Asymptotics of det\,{\boldmath$(I-K)$}\,---The nonperiodic case}\sp

The coefficients $a(\la)=a^+(\la)+a^-(\la),\ b(\la)=b^+(\la)+b^-(\la)$ of
the last section were in general given by integral formulas. They were
\bq a^+(\la)=-\int\,(H(s)\inv-1)\,{ds\over2\pi i},\label{ap}\eq
\bq b^+(\la)-a^+(\la)\,\log n=
\int\,(H_-(s)\inv-H_+(s)\inv)\,{1\over\Ga(s)}(\Ga(s)\,H_+(s))\,'\,{ds\over2\pi
i},
\label{bp}\eq
with the formulas for $a^-(\la),$ and $b^-(\la)$ obtained by replacing $H(s)$
by $H(1-s)$.
The integrations may be taken over $\Re\,s=\hf$ if $\la$ is small enough and,
as usual,
this is no loss of generality. To find $a$ and $b$ we integrate $a(\mu)$ and
$b(\mu)$,
respectively, with respect to $-d\mu/\mu$ over a path from 0 to $\la$.

Write
\[\ph(s):={\pi\over \sin\,\pi s}\,\int_{\Om}(-\om)^{s-1}\,d\rho(\om),\]
so that $H(s)=1-\la\,\ph(s)$. Making this replacement in the integrand in
(\ref{ap})
and integrating gives
\[-\int_0^{\la}\Big({1\ov 1-\mu\ph(s)}-1\Big){d\mu\ov\mu}=
-\int_0^{\la}{\ph(s)\ov 1-\mu\ph(s)}\,d\mu=\log\,(1-\la \ph(s)),\]
so the contribution to the coefficient of $\log\,t$ is
\[-\int\,\log\,(1-\la \ph(s)){ds\ov 2\pi i}\]
over $\Re\,s=\hf$. Replacing $H(s)$ by $H(1-s)$ gives the same contribution
since we may
make the substitution $s\ra 1-s$. Therefore
\bq a=-2\int\,\log\,(1-\la\ph(s)){ds\ov 2\pi i}.\label{arep}\eq
For general $\la$ the integration is to be on $\Re\,s=\sl$.

Now we go to (\ref{bp}), which may be written
\bq b^+(\la)-a^+(\la)\,\log n=\int\,{\Ga'(s)\ov\Ga(s)}\Big({1\ov
H(s)}-1\Big)\,{ds\over2\pi i}+
\int\,\Big({H_+'(s)\ov H_-(s)}-{H_+'(s)\ov H_+(s)}\Big)\,{ds\over2\pi
i}.\label{ints}\eq
By a computation like the earlier one we see that the first integral becomes
after
the $\mu$-integration
\[\int\,{\Ga'(s)\ov\Ga(s)}\,\log\,(1-\la\ph(s))\,{ds\ov 2\pi i}.\]
Then we replace $s$ by $1-s$, make the substitution $s\ra 1-s$, and add. We see
that the contribution of the first integral in (\ref{ints}) equals
\bq\int\,\Big({\Ga'(s)\ov\Ga(s)}+
{\Ga'(1-s)\ov\Ga(1-s)}\Big)\,\log\,(1-\la\ph(s))\,
{ds\ov 2\pi i}.\label{int1}\eq

Finally, we look at the second integral in (\ref{ints}), which equals
\[\int\,{H_+'(s)\ov H_-(s)}\,{ds\over2\pi i}
=\int\,{H_+'(s)\ov H_+(s)}H(s)\inv\,{ds\over2\pi
i}=\int\,(\log\,H_+)'(s)\,H(s)\inv\,
{ds\over2\pi i}.\]
Replacing $H(s)$ by $H(1-s)$ replaces $H_+(s)$ by $1/H_-(1-s)$, so after making
the
substitution $s\ra 1-s$ and adding we get
\bq\int\,(\log\,H_+\,H_-)'(s)\,H(s)\inv\,{ds\over2\pi i}.\label{int2}\eq
Recall that for $\Re\,s=\hf$
\[\log\,H_{\pm}(s)={\mp}\hf\,\log\,H(s)+\int\,{\log\,H(s')\ov
s'-s}\,{ds'\over2\pi i},\]
where the integral is the Hilbert transform, a principal value integral over
$\Re\,s'=\hf$.
So
\[\log\,H_+(s)\,H_-(s)=\int\,{\log\,H(s')\ov s'-s}\,{ds'\over\pi i}.\]
Since the Hilbert transform commutes with differentiation we get
\[(\log\,H_+\,H_-)'(s)={1\ov\pi i}\int\,{H'(s')\ov H(s')}\,{ds\ov s'-s},\]
and so (\ref{int2}) equals
\[-{1\ov 2\pi^2}\int\int\,{H'(s')\ov H(s')}\,H(s)\inv\,{ds'\ov s'-s}\,ds.\]
The $\mu$-integration gives
\[-\int_0^{\la}\,{-\mu\,\ph'(s')\ov 1-\mu\ph(s')}\,{1\ov
1-\mu\ph(s)}{d\mu\ov\mu}=
\ph'(s')\,\int_0^{\la}\,{1\ov (1-\mu\ph(s'))\,(1-\mu\ph(s))}\,d\mu\]
\[=\ph'(s')\,{1\ov\ph(s')-\ph(s)}\,\log\,{1-\la\ph(s)\ov1-\la\ph(s')},\]
and so the contribution of the second integral in (\ref{ints}) is
\bq -{1\ov
2\pi^2}\int\int\,{\ph'(s')\ov\ph(s')-
\ph(s)}\,\log\,{1-\la\,\ph(s)\ov1-\la\,\ph(s')}\,
{ds'\ov s'-s}\,ds.\label{int3}\eq
Thus $b-a\,\log n$ equals the sum of (\ref{int1}) and (\ref{int3}). As usual,
for general $\la$ the integrals are taken over $\Re\,s,\, s'=\sl$.\sp

\noi{\bf Remark}. The double integral (\ref{int3}) is exactly the constant in
the known asymptotics for the determinants of the truncated Wiener-Hopf
operators
associated with $\ph$ (specifically, $\ph(\hf+i\xi)$ is the Fourier transform
of the convolving kernel), and (\ref{arep}) is (minus twice) the leading
coefficient
in the asymptotics. One can see by the argument of Section 2 how both these
things
arise and conclude also that (\ref{int3}) equals $\,\det\,\cM\inv$. The extra
ingredient
here is therefore the integral (\ref{int1}).\sp

\setcounter{equation}{0}\renewcommand{\theequation}{6.\arabic{equation}}

\noi{\bf 6.} {\bf The case}\ {\boldmath$n=2$}\sp

In this case the only root is $\om=-1$ and we may take $c_{-1}$ equal to 1
since it
occurs only in the product $\la c_{-1}$. Thus the kernel of $K_0$ is
\[{e^{-2t(u+u\inv)}\over u+v}\]
and the equation (for either $q_k$) is
\bq q''(t)+t^{-1}q'(t)=8\,\sinh2q(t).\label{P3}\eq

We have in this case $h(s)=\sin\pi s-\pi\la$,
the zeros are given by
\[\al_0={1\ov\pi}\arcsin\pi\la={1\ov\pi i}\log(\pi i\la+\sqrt{1-\pi^2\la^2}),\
\ \
\al_1=1-\al_0,\]
and $\al_{k+2}=\al_k+2$. The square root is that branch which is positive for
$\la=0$
and the logarithm that branch which is 0 there.
 From this it is easy to see that the set (\ref{Lac})
consists of the rays $(-\iy,\,-1/\pi]$ and $[1/\pi,\,\iy)$ and $\La$, the
proposed region
of validity of our formulas, is the complex plane cut along these rays. If we
note
that the function $\pi i\la+\sqrt{1-\pi^2\la^2}$ map $\La$ onto the right
half-plane, we see that $|\Re\,\al_k(\la)-k|<\hf$ for all $\la\in\La$ and so
the
``extra'' condition on the $\al$ is satisfied. The range
of valididy is therefore all of $\La$. Using the formulas stated in the
introduction
we find that $\det\,(I-\la\,K_0)\sim b\,(t/2)^a$ with
\[a=\al_0^2+\al_0,\ \ \ b={G(\hf)\,G({3\ov2})\ov
G(\hf+\al_0)\,G({3\ov2}-\al_0)}=
{\Ga(\hf)\ov \Ga(\hf-\al_0)}\,{G(\hf)^2\ov G(\hf+\al_0)\,G({1\ov2}-\al_0)}.\]
For $\det\,(I+\la\,K_0)$ we replace $\la$ by $-\la$, which amounts to replacing
$\al_0$
by $-\al_0$. If we multiply the two results together we recover the asymptotics
for
$\det\,(I-\la^2\,K_0^2)$ determined in \cite{T} and \cite{BT}.

For $q_0$ we have the asymptotics $A\log(t/2)+\log B+o(1)$, where
\[A=2\,\al_0,\ \ \ B={\Ga(\hf-\al_0)\ov\Ga(\hf+\al_0)},\]
in agreement with \cite{MTW}. This is the solution of (\ref{P3}) which is
asymptotic
to $-2\la K_0(4t)$ as $t\ra\iy$, where this $K_0$ is the Bessel function.\sp

For $\la\not\in\La$ the asymptotics are different. For $\la>1/\pi$
$e^{q_0(t)}$ has an infinite sequence of zeros as $t\ra0$ and for $\la<-1/\pi$
it has
an infinite sequence of poles; this follows from the fact that as $t\ra0$ the
spectrum
of $K_0$ fills up the interval $[0,\,\pi]$. A heuristic derivation of the
asymptotics
for $\la$ on the cut is given in \cite{MTW}. In the next section we present a
similar derivation for some cases of $n=3$. \sp

\setcounter{equation}{0}\renewcommand{\theequation}{7.\arabic{equation}}

\noi{\bf 7.}\ {\boldmath$n=3$}\ {\bf and cylindrical Bullough-Dodd}\sp

The cylindrical Bullough-Dodd equation
\bq q''(t)+t^{-1}q'(t)=4 e^{2q} - 4 e^{-q}. \label{BD} \eq
arises in the special case of $n=3$ where $c_{\om}=-\om^3\,c_{\om\inv}$. Then
$q_1=0$, $q_2=-q_3$ and (\ref{BD}) is satisfied by $q=q_3$.
If we set $\zeta:=e^{2 \pi i/3}$ then $c_{\zeta}$ may be chosen arbitrarily. If
we  choose
it to be $\zeta(1-\zeta)$ then $c_{\zeta^2}=\zeta^2(1-\zeta^2),\ c_{-1}=0$ and
\[ h(s)=\sin\pi s + 2\pi\sqrt{3}\,\la \sin(\pi(s+2)/3). \]
Again $\la$ is the one free parameter. The zeros given by
\[\al_0={1\ov4}-{3\ov 2\pi}\arcsin\Big({1\over 2}+{\la\ov 2\la_c}\Big),\ \ \
\al_1=1,\ \ \
\al_2=2-\al_0,\]
where $\la_c=1/(2\sqrt{3}\pi)$.
Now $\La$ is the complement of the union of cuts
\[(-\iy,\,-3\la_c]\cup[\la_c,\,\iy).\]
For $\la\in\La$ the zeros satisfy
\[\Re\,\al_0\in(-\hf,\,1),\ \ \ \Re\,\al_1=1,\ \ \
\Re\,\al_2\in(1,\,{5\ov2}),\]
and so the extra condition is again automatically satisfied and our formulas
hold for
all $\la\in\La$. If we write
\bq q(t)= A \log({1\over t}) - \log B +o(1) \label{asy1}\eq
then the connection formulas give in this case
\[A=-2\al,\ \ \
B=3^{-A}\,{\Gamma\left({\al +2\ov 3}\right) \Gamma\left({2\al +1\ov
3}\right)\over
        \Gamma\left({1-\al \ov 3}\right) \Gamma\left({2-2\al\ov 3}\right)},\]
where we wrote $\al$ for $\al_0$. For large $t$
\[ q(t)\sim 6\la K_0\left(2\sqrt{3} t\right). \]\sp

\noi{\it Asymptotics at the critical value $\la_c$}\sp

This section and the following ones are heuristic. Using the differential
equation
(\ref{BD}) one can determine the correction terms to (\ref{asy1}):
\bq q(t)= A \log({1\over t}) + \log B + {B^2\over (1-A)^2}\, t^{2-2A}
-{4\over B (2+A)^2}\, t^{2+A} + {B^4\over 2 (1-A)^4}\, t^{4-4A} + \ldots
\label{asy2}\eq
This is valid  for $\la\in\La$.
To understand the higher order terms in more detail it is convenient to define
\[ w(t)=\exp\left(-q(t) \right) \]
where $w$ satisfies the equation
\bq w''={1\ov w} (w')^2-{1\ov t} w' + 4 w^2 -{4\ov w}. \label{wDE}\eq
The asymptotics we proved  become the statement
\[ w(t) = B t^A \left(1+o(1)\right). \]
Using (\ref{wDE}) to calculate the higher order terms in the small $t$
expansion for $w$ we find
\bq\begin{array}{rcl}\ds w(t)&\hspace{-1em}=\hspace{-1em}&\ds B t^A
\Big(1-{t^{2-2A}\over B^2(1-A)^2}+
{4B\over (2+A)^2}t^{2+A}+{12 B^2\over (2+A)^4} t^{4+2A}+ \cdots\\
&&\hspace{-1em}+\ds{(j+1)\, 2^j B^j \ov (2+A)^{2j}}\, t^{2j+jA} + \cdots\\
&&\hspace{-1em}+ \ds{24(A^2-2A-2)\ov (2+A)^2 (1-A)^2 (4-A)^2 B} t^{4-A}
+{4\over (1-A)^4 (4-A)^2  B^3} t^{6-3A} + \cdots\Big).\end{array}
\label{wAsy}\eq
In contrast to (\ref{asy2}) the terms $t^{2m-2mA}$ only appear for $m=1$ in the
above expansion.

As $\la$ varies from
0 to $\la_c$ $\al$ varies from 0 to $-\hf$, so $A$ varies from 0 to 1 and $B$
from
1 to $\iy$. Observe that the first two terms in (\ref{wAsy}) are of the same
order in $t$
as $t\ra0$ (and $A\ra1$) whereas the others are of lower order. This suggests
that
when $\la=\la_c$ we have $w(t)\sim t\Om_1$ as $t\ra0$, where
\bq
\Omega_1:=\lim_{\al\ra -\hf} B\left(1-{t^{2-2A}\ov B^2 (1-A)^2}\right)=
2\log(1/t)-{4\ov 3}\log 2 + 2\log 3 - 2\gamma.
\label{omega}\eq
We now use the differential equation (\ref{wDE}) to find the higher order
terms, which
are polynomial in  $t$ and $\Omega_1$. (The only property  of
$\Omega_1$ used  in the formal expansion is $d\Omega_1/dt = -2/t$.)
The expansion is
\bq
w(t) = t\Omega_1 +{4\ov 9} t^4 \left(\Omega_1^2+{4\ov 3} \Omega_1+{8\ov
9}\right)
+ {4\ov 2187} t^7 \left(81\Omega_1^3+216\Omega_1^2+240\Omega_1+80\right)
+O(t^{10} \Omega_1^4 ). \label{wAsy2}
\eq
Thus (as for the $n=2$ analogue \cite{MTW}) if one were to alter the
constant appearing in (\ref{omega}) then the solution of (\ref{wDE}) whose
asymptotics is
(\ref{wAsy2}) would not match onto the solution that approaches $1$ as
$t\ra\iy$.

These asymptotics at $\la=\la_c$ were checked by numerically solving
(\ref{wDE}) in both a foward and backward integration. There was agreement to
nine decimal places at $t=1/4$.\sp

\noi{\it Asymptotics at the critical value $-3\la_c$}\sp

We  proceed as above and examine all terms that would be of the same
order of magnitude as $\la\ra -3\la_c$, when $\al\ra1$ and $A\ra -2$. These are
the
terms of the geometric series, those involving the powers $t^{2j+jA}$. Summing
the
series we see that we must compute
\[
\lim_{\al\ra 1} {B \, t^A\ov \left(1- {2 B t^{2+A}/(2+A)^2}\right)^2}
={1\ov 2 t^2 \left( \log t - \log 3 + \gamma\right)^2}. \]
Defining
\[ \Omega_2 = \log t - \log 3 +\gamma \]
we thus see that at $\la=-3\la_c$,
\bq w(t) \sim {1\ov 2 t^2 \Omega_2^2 }. \label{wAsy4} \eq
To compute higher order terms it is convenient to look at $v(t) = 1/w(t)$.
Using the differential equation and only the property
$d\Om_2/dt =1/t$ of $\Om_2$ we find
\bq v(t)= 2 t^2 \Omega_2^2  + {t^8\ov 9}\left(8 \Omega_2^6-{32\ov 3}
\Omega_2^5+
{76\ov 9} \Omega_2^4 - {40\ov 9} \Omega_2^3+{40\ov 27} \Omega_2^2-{20\ov
81}\Omega_2
\right)+O(t^{14}\Om_2^{10}).\label{vAsy1} \eq \sp

\noi{\it Asymptotics for $\la>\la_c$}\sp

Think of $\la$ as being on the lower part of the cut $[\la_c,\iy)$. Then
\[\al=-\hf-{3\ov2}i\mu,\]  where
\[ \mu:={1\ov\pi}\arccosh\Big({1\ov 2}+{\la\ov 2\la_c}\Big),\ \ \ (\mu>0).\]
Thus $A=1+3i\mu$. Here again the first two terms in (\ref{wAsy}) are of the
same order
as $t\ra0$ whereas the others are of lower order, and we obtain
\[w(t)=Bt^A-{t^{1-2A}\ov B(1-A)^2}+O(t^4).\]
Substituting in the values of $A$ and $B$ in terms of $\mu$ we find
\[w(t) = \left({2 t\ov 3 \mu}\right) \Im\left[ e^{3i\mu\log(t/3)}
{ \Gamma(1/2-i\mu/2)\Gamma(-i\mu)\ov \Gamma(1/2+i\mu/2)\Gamma(i\mu)}\right]
+O(t^4).\]

If we had taken $\la$ to be on the upper part of the cut then we would have
replaced $\mu$ by $-\mu$. The result would have been precisely the same.\sp

\noi{\bf Remark}. In \cite{K} a method was described to find connection
formulas for
solutions of a class of equations including (\ref{BD}). Away from the critical
values the
short-range asymptotics stated there correspond to the first two terms in
(\ref{wAsy}).
As for the asymptotics at the critical values, our formulas agree with \cite{K}
at
$\la=-3\la_c$ but at $\la=\la_c$ we differ by a factor of 2.\sp\sp\sp
\pagebreak
\begin{center}{\bf Acknowledgemenmts}\end{center}

This work was supported in part by National Science Foundation Grants
DMS-9303413
(first author) and DMS-9424292 (second author). The authors also thank the
Volkswagen-
Stifftung for their support of the Research in Pairs program at Oberwolfach;
the first
results of the paper were obtained during the authors' visit under this
program.\sp


\begin{thebibliography}{4}

\bibitem{B} Barnes, E. W.: {\it The theory of the G-function}. Quart. J. Pure
and Appl. Math.
{\bf 31} 264--314 (1900)

\bibitem{BT} Basor, E. L., Tracy, C. A.: {\it Asymptotics of a tau function and
Toeplitz
determinants with singular generating functions}. Int. J. Mod. Phys. A {\bf 7}
83--107 (1992)

\bibitem{C} Cecotti, S., Fendley, P., Intriligator, K., Vafa, C.:
{\it A new supersymmetric index}, Nucl. Phys. {\bf B386}, 405--452 (1992)

\bibitem{GF} Gohberg, I. C., Feldman, I. A.: {\em Convolution equations and
projection methods
for their solution}, Transl. Math. Monogr. {\bf 41}, Providence:
Amer. Math. Soc.,  1974.

\bibitem{K} Kitaev, A. V.: {\it Method of isometric deformation for
``degenerate''
third Painlev\'e equation}. J. Soviet Math. {\bf 46}, 2077--2082 (1989)

\bibitem{MTW} McCoy, B. M., Tracy, C. A., Wu, T. T.: {\it Painlev\'e functions
of the
third kind}. J. Math. Phys. {\bf 18}, 1058--1092 (1977)

\bibitem{T} Tracy, C. A.: {\it Asymptotics of a $\tau$-function arising in the
two-dimensional Ising model}. Comm. Math. Phys. {\bf 142}, 297--311 (1991)

\bibitem{W1} Widom, H.: {Asymptotic behavior of block Toeplitz matrices and
determinants II}. Adv. in Math. {\bf 21},  1--29 (1976)

\bibitem{W2} Widom, H: {\it Some classes of solutions to the Toda lattice
hierarchy}.
Comm. Math. Phys. To appear. solv-int/9602001

\bibitem{Z} Zamolodchikov, Al. B.: {\it  Painlev{\'e} III and 2D polymers}.
Nucl. Phys. {\bf B432}[FS], 427--456 (1994)


\end{thebibliography}
\end{document}